\documentstyle[12pt,aps]{revtex}
\begin{document}
\newcommand{\be}{\begin{equation}}
\newcommand{\ee}{\end{equation}}
\newcommand{\ben}{\begin{eqnarray}}
\newcommand{\een}{\end{eqnarray}}
\newcommand{\bF}{\begin{figure}}
\newcommand{\eF}{\end{figure}}
\vskip 0.4in
\centerline{\Large\bf Quantum Mechanics As A Limiting Case}
\vskip 0.2in
\centerline{\Large\bf of Classical Mechanics}
\vskip 0.4in
\centerline{Partha Ghose}
\vskip 0.2in
\centerline{S. N. Bose National Centre for Basic Sciences}
\vskip 0.2in
\centerline{Block JD, Sector III, Salt Lake, Calcutta 700 091, India}
\vskip 0.6in
\abstract{In spite of its popularity, it has not been possible to vindicate 
the conventional wisdom that classical mechanics is a limiting case of quantum
mechanics. The purpose of the present paper is to offer an alternative point of
view in which quantum mechanics emerges as a limiting case of classical 
mechanics in which the classical system is decoupled from its environment.}

\vskip 0.2in

PACS no. 03.65.Bz

\newpage

\section{Introduction}

One of the most puzzling aspects of quantum mechanics is the quantum
measurement problem which lies at the heart of all its interpretations.
Without a measuring device that functions classically, there are no `events'
in quantum mechanics which postulates that the wave function contains
{\it complete} information of the system concerned and evolves linearly and
unitarily in accordance with the Schr\"{o}dinger equation. The system cannot
be said to `possess' physical properties like position and momentum
irrespective of the context in which such properties are measured. The
language of quantum mechanics is not that of realism.

According to
Bohr the classicality of a measuring device is
{\it fundamental} and cannot be {\it derived} from quantum theory. In other 
words,
the process of measurement cannot be analyzed within quantum theory itself.
A similar conclusion also follows from von Neumann's approach \cite{vN}.
In both these approaches the border
line between what is to be regarded as quantum or classical is, however,
arbitrary and mobile. This makes the theory intrinsically ill defined.

Some recent approaches have attempted
to {\it derive} the classical world from a quantum substratum by regarding
quantum systems as open. Their interaction with their `environment' can be
shown to lead to effective {\it decoherence} and the emergence of quasi-
classical behaviour
\cite{Joos}, \cite{Zurek1}. However, the very concepts of a `system' and its
`environment' already presuppose a clear cut division between them which, as
we have remarked, is mobile and
ambiguous in quantum mechanics. Moreover, the reduced density matrix of the
`system' evolves to a diagonal form only in the pointer basis and not in the
other possible bases one could have chosen.
This shows that this
approach does not lead to a real solution of the measurement problem, as
claimed by Zurek \cite{Zurek2}, though it is an important development that
sheds new light on the emergence of quasi-classical
behaviour from a quantum susbstratum.

The de Broglie-Bohm approach \cite{BH}, on the other hand, does not accept the
wave function description as complete.
Completeness is achieved by introducing the position of the particle as an
additional variable (the so-called `hidden variable') with
an {\it ontological} status. The wave function at a
point is no longer just the probability amplitude that a particle will be
{\it found} there if a measurement were to be made, but the probability
amplitude that a particle {\it is} there even if no measurement is made.
It is a realistic description, and measurements are reduced to ordinary
interactions and lose their mystique. Also, the classical limit is much
better defined in this approach through the `quantum potential' than in
the conventional approach. As a result, however, a new problem is unearthed,
namely, it becomes quite clear that classical theory admits ensembles of a
more general kind than can be reached from standard quantum ensembles.
The two theories are really disparate while having a common domain of
application \cite{Holland}.

Thus, although it is tacitly assumed by most physicists
that classical physics is a limiting case of quantum
theory, it is by no means so. Most physicists would, of course,
scoff at the suggestion that the situation may really be the other way round,
namely, that quantum mechanics is contained in a certain sense in
classical theory. This seems
impossible because quantum mechanics includes totally new elements like
$\hbar$ and the uncertainty relations and the host of new results that follow
from them. Yet, a little reflection shows that if true classical behaviour
of a system were really to result from a quantum substratum through some
process analogous to `decoherence', its quantum behaviour ought also to
emerge on isolating it sufficiently well
from its environment, i.e., by a process which is
the `reverse of decoherence'. In practice, of course, it would be impossible
to reverse decoherence once it occurs for a system.
Nevertheless, it should still be possible to prepare a system sufficiently
well isolated from its environment so that its quantum behaviour can be
observed. If this were not possible, it would have been impossible
ever to observe the quantum features of any
system. 

So, let us examine what the opposite point of view implies,
namely that classical theory is more fundamental than quantum
theory (in a sense to be defined more precisely). This would, in 
fact, be consistent with Bohr's
position that the classicality of measuring devices is fundamental
(nonderivable), leading to his preferred solution to the
quantum measurement problem. At the same time, the approach of de Broglie and
Bohm {\it coupled with the notion of decoherence as an environmental
effect that can be switched on} would fall into place, but the
non-realist Copenhagen interpretation would have to be abandoned.

\section{The Hamilton-Jacobi Theory}

Our starting point is the non-relativistic Hamilton-Jacobi equation

\be
\partial S_{cl}/\partial t +  \frac{(\,\nabla\,S_{cl})^2}{2 m}\,+
\,V(x) = 0
\label{eq:a}
 \ee
for the action $S_{cl}$ of a classical paticle in an external potential $V$,
together with the definition of the momentum

\be
{\bf p}  = m\,\frac{d {\bf x}}{d t}
= {\bf \nabla} S_{cl}
\label{eq:b}
\ee
and the continuity equation

\be
\frac{\partial \rho_{cl} ( {\bf x}, t )}{\partial t} +
{\bf \nabla}\,.\, (\,\rho_{cl}\, \frac{{\bf \nabla}\,S_{cl} }{m}) = 0
\label{eq:c}
\ee
for the position distribution function $\rho_{cl} ( {\bf x}, t )$ of the 
ensemble of trajectories
generated by solutions of equation (\ref{eq:a}) with different initial
conditions (position or momentum).
Suppose we introduce a complex wave function

\be
\psi_{cl}\,(\,{\bf x}\,,\,t\,) = R_{cl}\,(\,{\bf x}\,,\,t\,)\,
exp\,(\,{\frac{i}{\hbar}\,S_{cl}})
\label{eq:d}
\ee
into the formalism by means of the equation

\be
\rho_{cl}\,(\,{\bf x}\,,\,t\,) = \psi_{cl}^*\,\psi_{cl} = R_{cl}^2\,.
\label{eq:e}
\ee
What is the equation that this wave function must satisfy such that the
fundamental equations (\ref{eq:a}) and (\ref{eq:c}) remain unmodified?
The answer turns out to be the modified Schr\"{o}dinger equation
\cite{Holland}

\be
i\hbar\,\frac{\partial \psi_{cl}}{\partial t} = \left(-\frac{\hbar^2}{2 m}\,
\nabla^2  + V(x)\right)\,\psi_{cl} - Q_{cl}\,\psi_{cl}
\label{eq:f}
\ee
where

\be
Q_{cl} = - \frac{\hbar^2}{2 m}\,\frac{\nabla^2 R_{cl}}{R_{cl}}
\label{eq:g}
\ee
Thus, a system can behave classically in spite of it having an associated
wave function that satisfies this modified Schr\"{o}dinger equation.

Notice that the last term in this equation is nonlinear in
$\vert\psi_{cl}\vert$, and is {\it uniquely} determined by the requirement that
all quantum mechanical effects such as superposition, entanglement and
nonlocality be eliminated. It is therefore to be sharply
distinguished from certain other types of nonlinear terms that have been
considered in constructing nonlinear versions of quantum mechanics
\cite{Weinberg}.
An unacceptable consequence of such nonlinear terms (which are, unlike
$Q_{cl}$, bilinear
in the wave function) is that superluminal signalling using quantum
entanglement becomes possible in such theories \cite{Gisin}. Since
$Q_{cl}$ eliminates quantum superposition and entanglement, it cannot
imply any such possibility. Usual action-at-a-distance is, of course,
implicit in non-relativistic mechanics, and can be eliminated in a Lorentz
invariant version of the theory, as we will see later.

Deterministic nonlinear terms with arbitrary parameters have also been
introduced in the Schr\"{o}dinger equation to bring about
collapse of quantum correlations \cite{GRW} for isolated macroscopic
systems. Such terms also imply superluminal signals via quantum entanglement. 
The term 
$Q_{cl}$ is different from such terms as well in that it
has no arbitrary parameters in it and eliminates
quantum correlations for all systems deterministically, irrespective
of their size.
 
{\it Most importantly, it is clear from the above analysis that none of the 
other types of nonlinearity can guarantee 
strictly classical behaviour described by equations (\ref{eq:a}) and (\ref{eq:c}).}

Let us now consider the classical version of the density matrix which must
be of the form

\ben
\rho_{cl} ( x, x^{'}, t ) &=& R_{cl} ( x, t )
exp\biggl ( \frac{i}{\hbar} S_{cl} ( x, t )\biggr) R_{cl} ( x^{'}, t )
exp\biggl ( \frac{i}{\hbar} S_{cl} ( x^{'}, t )\biggr)\\
&=& R^2 ( x, t ) \delta^{3} ( x - x^{'} )
\een
in order to satisfy the Pauli master equation. The absence of off-diagonal terms is a consequence of the absence of
quantum correlations between spatially separated points. This implies
that the classical wave function can be written as

\be
\psi_{cl} ( x, t ) = \frac{1}{\sqrt{\pi^3}}\lim_{\epsilon
\rightarrow 0} \sqrt{\frac{\epsilon}{(x -x(t) )^2 + \epsilon^2}}\,
exp (\frac{i}{\hbar} S_{cl} )\,.
\ee
Such a function has only point
support on the particle trajectory $x = x(t)$ determined by equation
(\ref{eq:b}). It can also be written as a linear superposition of the
delta function and its derivatives \cite{Gelfand}. All this ensures 
a classical phase space.

The wave function $\psi_{cl}$ is therefore entirely dispensable and
``sterile'' as long as we
consider strictly classical systems. Conceptually, however, it acquires
a special significance in considering the transition between quantum
and classical mechanics, as we will see.

The wave function $\psi$  of a quantum mechanical
system, on the other hand, must of course satisfy the
Schr\"{o}dinger equation

\be
i\,\hbar\,\frac{\partial \psi}{\partial t} = -\frac{\hbar^2}{2m}\,\nabla^2 \psi
+ V\,\psi\,\,.
\label{eq:i}
\ee
Using a polar representation similar to (\ref{eq:d}) for $\psi$ in this
equation and separating
the real and imaginary parts, one can now derive the {\it modified}
Hamilton-Jacobi equation

\be
\partial S/\partial t + \frac{(\nabla S)^2}{2m} + Q + V = 0
\label{eq:j}
\ee
for the phase $S$ of the wave function, where $Q$ is given by

\be
Q = - \frac{\hbar^2}{2 m}\,\frac{\nabla^2 R}{R}\,,
\label{eq:g2}
\ee
and the continuity equation

\be
\frac{\partial \rho ( {\bf x}, t )}{\partial t} +
{\bf \nabla}\,.\, (\,\rho\, \frac{{\bf \nabla}\,S }{m}) = 0
\label{eq:c2}
\ee
These differential equations ((\ref{eq:j}) and (\ref{eq:c2})) now become
coupled differential equations which determine $S$ and $\rho = R^2$.
Note that {\it the phase $S$ of a quantum mechanical system
satisfies a modified
Hamilton-Jacobi equation with an additional potential $Q$ called the
``quantum potential''.} Its properties are therefore different from
those of the
classical action $S_{cl}$ which satisfies equation (\ref{eq:a}) .
Applying the operator ${\bf \nabla}$ on equation
(\ref{eq:j}) and using the definition of the momentum (\ref{eq:b}), one
obtains the equation of motion

\be
 \frac{d \bf p}{d t} = m\,\frac{d^2\,{\bf x}}{d t^2} = -\, {\bf \nabla}\,(\,V + Q)
\label{eq:k}
\ee
for the quantum particle. Integrating this equation or, equivalently  
equation (\ref{eq:b}), one obtains the Bohmian trajectories $x(t)$ of the 
particle corresponding to different initial positions.
The departure from the classical Newtonian
equation due to the presence of the ``quantum potential'' $Q$ gives
rise to all the
quantum mechanical phenomena such as the existence of discrete stationary
states, interference phenomena, nonlocality and so on. This agreement with quantum mechanics is achieved by requiring that the initial
distribution $P$ of the particle is given
by $R^2 (\,x(t)\,, 0\,)$. The continuity
equation (\ref{eq:c2}) then guarantees that it will agree with $R^2$ at
all future times.
This guarantees that the  
averages of all dynamical
variables of the particle taken over a Gibbs ensemble of its trajectories 
will always agree with
the expectation
values of the corresponding hermitian operators in standard
quantum mechanics.
This is essentially the de Broglie-Bohm quantum theory of
motion. For further details about this theory and its relationship
with standard
quantum mechanics, the reader is referred to the comprehensive book by
Holland \cite{Holland} and the one by Bohm and Hiley \cite{BH}.

Now, let us for the time being assume that quantum mechanics is the more
fundamental theory from which classical mechanics follows in some limit.
Consider a quantum mechanical system interacting with its
environment. It evolves according to the Schr\"{o}dinger equation

\be
i\,\hbar\,\frac{\partial \psi}{\partial t} = \left(-\frac{\hbar^2}{2 m}\,
\nabla^2  + V(x)  + W \,\right)\,\psi
\label{eq:h}
\ee
where $W$ is the potential due to the environment experienced
by the system. For a complex enough environment such as a heat bath,
the density matrix of the system in the position representation
quickly evolves to a diagonal form. In a special model in which a
particle interacts only with the thermal excitations of a scalar field
in the high temperature limit, the density matrix evolves
according to the {\it master equation} \cite{Zurek}

\be
\frac{d \rho}{d t} = - \gamma ( x - x^{'} )
( \partial_{x} - \partial_{x^{'}} ) \rho -
\frac{2 m \gamma k_{B} T}{\hbar^2} ( x - x^{'})^2 \rho
\ee
where $\gamma$ is the relaxation rate, $k_{B}$ is the Boltzmann
constant and $T$ the temperature of the field.
It follows from this equation that quantum coherence falls off at
large separations as the square of $\Delta x = (x - x^{'})$. The
decoherence time scale is given by

\be
\tau_{D} \approx \tau_{R} \frac{\hbar^2}{2 m k_{B} (\Delta x)^2} =
\gamma^{-1} \biggl( \frac{\lambda_{T}}{\Delta x} \biggr)^2
\label{eq:X}
\ee
where $\lambda_{T} = \hbar/\sqrt{2 m k_{B} T}$ is the thermal de Broglie wavelength and $\tau_{R} = \gamma^{-1}$. For a macroscopic object
of mass $m = 1$ g at room temperature ( $T = 300 K$) and separation
$\Delta x = 1$ cm, the ratio $\tau_{D}/\tau_{R} = 10^{- 40}$ ! Thus,
even if the relaxation time was of the order of the age of the universe,
$\tau_{R} \simeq 10^{17}$ sec, quantum coherence would be destroyed in
$\tau_{D} \simeq 10^{- 23}$ sec. For an electron, however, $\tau_{D}$
can be much more than $\tau_{R}$ on atomic and larger scales.

However, the diagonal matrix does not
become diagonal in, for example, the momentum representation, showing that
coherence has not really been destroyed. The
FAPP diagonal density matrix does not therefore represent a {\it proper}
mixture of mutually exclusive alternatives, the classical limit is not
really achieved and the measurement problem remains \cite{Pearle}.

This is not hard to understand once one realizes that a
true classical system must be governed by a
Schr\"{o}dinger equation that is {\it modified} by the addition of a unique term
that is nonlinear in $\vert \psi \vert$ (equation (\ref{eq:f})), and that
{\it such a nonlinear term cannot arise from unitary Schr\"{o}dinger
evolution.}  On the contrary, it is not unnatural to expect a linear
equation of the Schr\"{o}dinger type to be the limiting case
of a nonlinear equation like equation (\ref{eq:f}). It is therefore
tempting to interpret the last term in equation (\ref{eq:f})
as an `effective' potential that represents the coupling of the classical
system to its environment. It is important to bear in mind that in
such an interpretation, the potential $Q_{cl}$ must obviously be
regarded as {\it fundamentally given} and {\it not derivable from a
quantum mechanical substratum}, being uniquely
and solely determined by the requirement of classicality, as shown above.

Let us now consider a quantum system which is inserted into
a thermal bath at time $t = 0$. If it is to evolve into a genuinely classical
system after a sufficient lapse of time $\Delta t$, its wave function
$\psi$ must satisfy the equation of motion

\ben
i\,\hbar\,\frac{\partial \psi}{\partial t} = \left(-\frac{\hbar^2}{2 m}\,
\nabla^2  + V(x)  - \lambda (t) Q_{cl} \,\right)\,\psi
\label{eq:h1}
\een
where $\lambda (0) = 0$ in the purely quantum limit and
$\lambda (\Delta t) = 1$ in the purely
classical limit. (Here $\Delta t
\gg\tau_{D}$ where $\tau_{D}$ is typically given by $\gamma^{-1}
(\lambda_{T}/\Delta x )^2$ (\ref{eq:X}).) Thus,
for example, if $ =\lambda (t) = 1 - exp ( -t /\tau_{D} )$,
a macroscopic system would very rapidly behave like a true classical
system at
sufficiently high temperatures, whereas a mesoscopic system
would behave neither fully like a classical system nor fully like a
quantum mechanical system at appropriate temperatures for a much
longer time. What happens is that the reduced density operator of the 
system evolves
according to the equation

\ben
\rho (x, x^{'}, \Delta t ) &=& exp (- i \int_{0}^{\Delta t}\lambda Q_{cl} 
d t/\hbar) 
\rho (x, x^{'},  0)
exp ( i \int_{0}^{\Delta t}\lambda Q_{cl} d t/\hbar )\\
&=& R^2 ( x, \Delta t ) \delta^{3} ( x - x^{'} )
\een
during the time interval $\Delta t$ during which the nonlinear interaction
$\lambda Q_{cl}$ completely destroys
all superpositions, so that at the end of this time interval the system
is fully classical and the equation
for the density operator reduces to the Pauli master equation for a
classical system.

A variety of functions $\lambda (t)$ would satisfy the
requirement $\lambda=0$ and $\lambda=1$. This is not surprising and is probably a reflection of the diverse ways in which different systems decohere in different environments.

It is clear that a system must be extremely well isolated ($\lambda = 0$) for it to behave
quantum mechanically. Such a system, however,
would inherit only a de Broglie-Bohm ontological and causal interpretation,
not an interpretation of the Copenhagen type. The practical difficulty is
that once a quantum system and its environment get coupled, it becomes FAPP
impossible to decouple them in finite time because
of the extremely large number of degrees of freedom of the environment.
However, we know from experience that it is possible
to {\it create} quantum states in the laboratory that are very well
isolated from their environment. Microscopic quantum systems are, of course, routinely
created in the laboratory (such as single atoms, single electrons, single
photons, etc.,) and considerable
effort is being made to create isolated macroscopic systems that would show
quantum coherence, and there is already some evidence of the existence of
mesoscopic `cat states' which decohere when appropriate radiation is introduced into the cavity \cite{Brune}.

Equation (\ref{eq:h1}) is a totally new equation that correctly bridges the gap 
between the quantum and the classical worlds. It should form a sound starting point for studying systems, parametrized by $\lambda(t)$, that lie anywhere in the continuous spectrum stretching between the quantum and classical limits. 

Notice that if one defines the momentum by the relation $\pi = \nabla S
- \int \nabla Q dt$, the equation of motion can be written in the
classical form

\be
\frac{d \pi}{d t} =- \nabla V\,.
\ee
This shows that it is $\pi$ which is conserved in the absence of any external
potential and not the particle momentum $p$. This is obviously due to the
existence of the quantum potential.

A look at the modified Hamilton-Jacobi equation (\ref{eq:j}) also shows that the quantity conserved by it is not the classical energy but this energy plus the quantum potential. Also notice that the equation of motion (\ref{eq:k}) implies that a quantum mechanical particle is not free even in the absence of an external potential. It is obvious therefore that the interaction of the corresponding classical system with its environment must serve to cancel this purely quantum force and restore the classical laws of motion. Once the form of the classical Hamilton-Jacobi equation is restored, conservation of energy is mathematically inevitable.

Notice that the additional interaction of a classical system with its environment in the form of the effective potential $Q_{cl}$ becomes manifest only when the Hamilton-Jacobi equation is recast in terms of the classical wave function (equations (\ref{eq:f}) and (\ref{eq:g})). This is why the Hamilton-Jacobi equation can be written without ever knowing about this interaction. The wave function approach reveals what lies hidden and sterile in the traditional classical approach. This is a significant new insight offered by the wave function approach.

It is important to point out a fundamental difference between the two
potentials $V(x)$ and $Q$ in (\ref{eq:j}). $V(x)$ is a given
external potential whereas $Q$ is not so---it depends
on the modulus of the wave function of the system,
and is therefore nonlocal in character.

This leads to a fundamental difference of the approach advocated in
this paper from the conventional de Broglie-Bohm theory in which
quantum mechanics rather than classical mechanics is regarded as being more
fundamental. In the de Broglie-Bohm theory the quantum potential
must necessarily vanish in the classical limit, and the quantum system
{\it appears} to behave classically. On the other hand, in the present
approach there is no need for the quantum
potential to vanish in the classical limit---only its effects must be
completely {\it cancelled} by  {\it nonlinear} environment-induced decoherence of a very special type.
Furthermore, besides the wave function, de Broglie and Bohm must also
introduce the position of the particle as an additional variable to complete the
description of the system. If classical mechanics happens to be more
fundamental than quantum mechanics, there is no need to do this as the
position and trajectory are already present in the fundamental
description. It is , in fact,  the wave function that acquires a
subsidiary role in this approach.
It is interesting that some circumstantial evidence already seems
to exist indicating that the position of
a quantum system plays a more fundamental role than its wave function \cite{Home}.

There is therefore a fairly strong case in favour of the
possibility that quantum theory might be the
limiting case of classical mechanics in which the interaction of the
system with its environment (nonlinear in $\vert \psi\vert$) is completely
switched off. It is difficult to see how such a situation can be accommodated
within the standard Copenhagen philosophy. The wave function also acquires
a new significance---it is sterile and dispensable in the classical limit
but becomes potent and indispensable in the quantum limit.

\section{The Klein-Gordon Equation}

Let the Hamilton-Jacobi equation for free relativistic classical particles
be

\be
\frac{\partial S_{cl}}{\partial t} + \sqrt{ (\partial_i S_{cl})^2\,c^2 + m_0^2\,c^4}
= 0\,.
\label{eq:n}
\ee
Then, using the relation $p_\mu = - \partial_\mu S_{cl} = m_0\,u_\mu$ where
$u_\mu = d\,x_\mu/d\,\tau$ with $\tau = \gamma^{-1}\,t$, $\gamma^{- 1}
= \sqrt{1 - v^2/c^2}, v_i = d\,x_i/d\,t$, the particle equation of motion is
postulated to be

\be
m_0\,\frac{d u_\mu}{d \tau} = 0 = \frac{d\,p_\mu}{d\,\tau} \,.
\label{eq:o}
\ee
It is quite easy to show that the classical
equations (\ref{eq:n}) and (\ref{eq:c}) continue to hold if one describes the
system in terms of a complex wave function
$\psi_{cl} = R_{cl}\, exp\,(\,\frac{i}{\hbar} S_{cl}\,)$ that satisfies the modified
Klein-Gordon equation

\be
\left(\,\Box + \frac{m_0^2\, c^2}{\hbar^2} - \frac{ Q_{cl}}{\hbar^2} \right)
\,\psi_{cl} = 0
\label{eq:p}
\ee
with

\be
Q_{cl} = \hbar^2 \frac{\Box R_{cl}}{R_{cl}}\,.
\label{eq:q}
\ee
As in the non-relativistic case, $ Q_{cl}$ may be interpreted as an effective
potential in which the system
finds itself when described in terms of the wave function $\psi_{cl}$.
If this potential goes to zero in some limit, one obtains the free Klein-Gordon
equation which is the quantum limit.

On the other hand, using $\psi = R\, exp\,(\,\frac{i}{\hbar} S\,)$ in the
Klein-Gordon equation
and separating the real and imaginary parts, one obtainds respectively
the equation

\be
\frac{1}{c^2}\,\left(\frac{\partial S}{\partial t}\right)^2
- \left( \partial_i S\right)^2 - m_0^2\, c^2 - Q = 0
\label{eq:r1}
\ee
which is equivalent to the modified Hamilton-Jacobi equation

\be
\left(\frac{\partial S}{\partial t}\right)
+ \sqrt{ \left( \partial_i S\right)^2\,c^2 + m_0^2\, c^4 + c^2\,Q}
= 0
\label{eq:r}
\ee
and the continuity equation

\be
\partial^\mu\,(\,R^2 \partial_\mu\,S\,) = 0\,.
\ee
One can then identify the four-current as $j_\mu = - R^2 \partial_\mu S$ so
that $\rho = j_0 =  R^2 E/c$ which is not positive definite because $E$ can be
either positive or negative, and therefore, as is well known,
it is not possible to interpret it as a probability density.

Nevertheless, let us note in passing that, if use is made of the definition
$p_\mu = - \partial_\mu\,S$ of the particle four-momentum, (\ref{eq:r1})
implies

\be
p_\mu\,p^\mu = m_0\,c^2 + Q
\label{eq:t}
\ee
and $p_\mu = M_0\,u_\mu$ where $M_0 = m_0\,\sqrt{1 +
Q/m_0^2\,c^2}$. Thus, the quantum potential $Q$ acts on the particles and
contributes to their energy-momentum so that they are off their mass-shell.
\footnote{The author is grateful to E. C. G. Sudarshan for drawing his
attention to this important point.}
Applying the operator $\partial_\mu$ on equation (\ref{eq:r1}), we get the
equation of motion

\be
\frac{d\,p_\mu}{d\,\tau} = \frac{\partial_\mu\,Q}{2\,M_0}
\ee
which has the correct non-relativistic limit. The equation for
the acceleration of the particle is therefore given by \cite{Holland}

\be
\frac{d\,u_\mu}{d\,\tau} = \frac{1}{2\,m_0^2}\,
(\,c^2\,g_{\mu\nu} - u_\mu\,u_\nu\,)\,\partial^\nu\,log\,(1 + \frac{Q}{m_0^2\,c^2}\,)\,.
\ee
If, on the other hand, one uses the modified Klein-Gordan equation
(\ref{eq:p}) and the corresponding Hamilton-Jacobi equation (\ref{eq:n}),
the particles are on their mass-shell and the free particle classical
equation (\ref{eq:o}) is satisfied.

\section{Relativistic spin 1/2 particles}

Let us now examine the Dirac equation for relativistic spin $1/2$ particles,

\be
(\,i \hbar \gamma_\mu \partial^\mu + m_0\, c )\,\psi = 0.
\label{eq:D1}
\ee
Let us write the components of the wave function
$\psi$ as $\psi^a = R\,\theta^a\,exp\,\,(\,\frac{i}{\hbar}\,S^a)$,
$\theta^a$ being a spinor component. It is not straightforward
here to separate the real and imaginary parts as in the previous cases.
One must therefore follow a different method for relativistic fermions.

It is well known that every component $\psi^a$ of the Dirac wave function
satisfies the Klein-Gordan equation. It follows therefore, by putting
$\psi^a = R \theta^a\,exp\,(\,i\,S^a/\hbar\,)$, that $S^a$ must satisfy the modified
Hamilton-Jacobi equation

\be
\partial_\mu\,S^a\,\partial^\mu\,S^a - m_0^2\, c^2 - Q^a = 0\,.
\label{eq:D2}
\ee
where $Q^a = \hbar^2\,\Box\,R\,\theta^a/R\,\theta^a$.
Summing over $a$, we get

\be
\sum_a\,\partial_\mu\,S^a\,\partial^\mu\,S^a - 4\,m_0^2\, c^2 -
\sum_a\,Q^a = 0\,.
\label{eq:D3}
\ee
Defining

\ben
\partial_\mu\,S\,\partial^\mu\,S &=& \frac{1}{4}\,
\sum_a\,\partial_\mu\,S^a\,\partial^\mu\,S^a\\
Q &=& \frac{1}{4}\,\sum_a\,Q^a\,,
\label{eq:D4}
\een
we have

\be
\partial_\mu\,S\,\partial^\mu\,S - m_0^2\, c^2 -
Q = 0\,.
\label{eq:D5}
\ee
Then, defining the particle four-momentum by $p_\mu = - \partial_\mu\,S$,
one has $p_\mu\,p^\mu = m_0^2\,c^2 + Q$. Therefore, one has the equation of
motion

\be
\frac{d\,p_\mu}{d\,\tau} =
\frac{\partial_\mu\,Q}{2\, M_0}\,.
\label{eq:D6}
\ee
The Bohmian 3-velocity of these particles is defined by the relation

\be
v_i = \gamma^{- 1}\,u_i = c\,\frac{u_i}{u_0} = c\,\frac{j_i}{j_0}
= c\,\frac{\psi^{\dagger}\,\alpha_i\,\psi}{\psi^{\dagger}\,\psi}\,.
\label{eq:D7}
\ee
Then, it follows that

\be
u_\mu = \gamma\,v_\mu = \gamma\,c\,\frac{j_\mu}{\rho}
\label{eq:D8}
\ee
where $\rho = \psi^{\dagger}\,\psi$. This relation is satisfied because
$j_\mu\,j^\mu = \rho^2/\gamma^2$ if (\ref{eq:D7}) holds.

As we have seen, for a classical theory of spinless particles,
the correct equation for the associated wave function is
the modified Klein-Gordon equation (\ref{eq:p}). Let the corresponding
modified wave equation for classical spin $1/2$ particles be of the form

\be
\left(\,i\, \hbar\, \gamma_\mu\, D^\mu + m_0\, c \,\right)\,\psi_{cl} = 0
\label{eq:D9}
\ee
where $D^\mu = \partial^\mu + (i/\hbar)\,Q^\mu$. Then we have

\be
(\,D_\mu\,D^\mu + \frac{m_0^2\,c^2}{\hbar^2}\,)\,\psi_{cl}^a = 0\,.
\label{eq:D10}
\ee
Writing $\psi_{cl}^a = R_{cl}\,\theta^a\,exp\,(\frac{i}{\hbar}\,S_{cl}^a )$, one obtains

\be
\partial_\mu\,S_{cl}^a\,\partial^\mu\,S_{cl}^a - m_0^2\, c^2 - Q_{cl}^a + Q_\mu\,Q^\mu
- 2\,Q_\mu\,\partial^\mu\,S_{cl}^a = 0
\label{eq:D11}
\ee
where 

\be
Q^{a}_{cl}= \frac{\hbar^2 \Box R_{cl} \theta^{a}}{R_{cl} \theta^{a}}.
\ee 
Define a diagonal matrix
$B_\mu^{a\,b} \equiv \partial_\mu\,S_{cl}^a\,\delta^{a\,b}$ such that

\be
\frac{1}{2}\,Tr B_\mu = \frac{1}{2}\,\sum_a\,\partial_\mu\,S_{cl}^a
\equiv \partial_\mu\,S_{cl}\,.
\ee
Then

\ben
\partial_\mu\,S_{cl}\,\partial^\mu\,S_{cl} &=& \frac{1}{4}\,Tr\,B_\mu\,Tr\,B^\mu =
\frac{1}{4}\,Tr\,(\,B_\mu\,B^\mu\,)\\&=& \frac{1}{4}\,\sum_a
\partial_\mu S_{cl}^a\,\partial^\mu S_{cl}^a\,.
\een
Therefore, taking equation (\ref{eq:D11}) and summing over $a$, we have

\be
\partial_\mu\,S_{cl}\,\partial^\mu\,S_{cl} - m_0^2\, c^2 -
Q_{cl} + Q_\mu\,Q^\mu - Q_\mu\,\partial^\mu\,S_{cl} = 0
\label{eq:D12}
\ee
where 

\be
Q_{cl} = \frac{1}{4} \sum_a Q^a_{cl}\,.
\ee
In order that the classical free particle equation is satisfied, the effects
of the quantum potential must be cancelled by this additional interaction,
and one must have

\be
Q_\mu\,(\,Q^\mu - \partial^\mu\,S_{cl}\,) = Q_{cl}\,.
\ee
A solution is given by

\ben
p_\mu &=& - \partial_\mu\,S_{cl} = m_0\,u_\mu\,,\\
Q_\mu &=& \alpha\,m_0\,u_\mu
\een
with

\be
\alpha = \frac{1}{2} \pm \frac{1}{2}\,\sqrt{1 + 4\,Q_{cl}/m_0^2\,c^2}\,.
\ee

\section{Relativistic spin 0 and spin 1 particles}

It has been shown \cite{Ghose} that a consistent relativistic quantum
mechanics of spin 0 and spin 1 bosons can be developed using the
Kemmer equation \cite{Kemmer}

\be
(\,i\,\hbar\,\beta_\mu\,\partial^\mu + m_0\,c\,)\,\psi = 0\,
\label{eq:1}
\ee
where the matrices $\beta$ satisfy the algebra

\be
\beta_{\mu}\,\beta_{\nu}\,\beta_{\lambda} + \beta_{\lambda}\,\beta_{\nu}\,
\beta{\mu} = \beta_{\mu}\,g_{\nu \lambda} + \beta_{\lambda}\,g_{\nu \mu}\,.
\label{eq:2}
\ee
The $5\times 5$ dimensional representation of these matrices describes spin 0
bosons and the $10 \times 10$ dimensional representation describes spin 1
bosons. Multiplying (\ref{eq:1}) by $\beta_0$, one obtains the
Schr\"{o}dinger form of the equation

\be
i\,\hbar\,\frac{\partial \psi}{d t} = [\,- i\,\hbar\,c\, \tilde{\beta}_i\,
\partial_i - m_0\,c^2\,\beta_0\,]\,\psi
\label{eq:3}
\ee
where $\tilde{\beta}_i \equiv \beta_0\,\beta_i - \beta_i\,\beta_0$. Multiplying
(\ref{eq:1}) by $1- \beta_0^2$, one obtains the first class constraint

\be
i\,\hbar\,\beta_i\,\beta_0^2\,\partial_i\,\psi = -m_0\,c\,(\,1 - \beta_0^2\,)
\,\psi.
\label{eq:4}
\ee
The reader is referred to Ref. \cite{Ghose} for further discussions regarding
the significance of this constraint.

If one multiplies equation (\ref{eq:3}) by $\psi^{\dagger}$ from the left,
its hermitian conjugate by $\psi$ from the right and adds the resultant
equations, one obtains the continuity equation

\be
\frac{\partial\,( \psi^{\dagger}\,\psi )}{\partial t} + \partial_i\,
\psi^{\dagger}\,\tilde{\beta}_i\,\psi = 0\,.
\ee
This can be written in the form

\be
\partial^\mu\,\Theta_{\mu 0} = 0
\ee
where $\Theta_{\mu \nu}$ is the symmetric energy-momentum tensor with
$\Theta_{0 0} = - m_0\,c^2\,\psi^{\dagger}\,\psi < 0$. Thus, one can define
a wavefunction $\phi = \sqrt{m_0\,c^2/E}\,\psi$ (with $E =- \int\,\Theta_{0 0}
\,dV$ ) such that
$\phi^{\dagger}\,\phi$ is non-negative and normalized and can be
interpreted as a probability density. The conserved probability current
density is $s_\mu = - \Theta_{\mu 0}/E = (\,\phi^{\dagger}\,\phi,
- \phi^{\dagger}\,\tilde{\beta}_i\,\phi )$ \cite{Ghose}.

Notice that according to the equation of motion (\ref{eq:3}), the velocity
operator for massive bosons is $c\,\tilde{\beta}_i$, so that the Bohmian
3-velocity can be defined by

\be
v_i = \gamma^{- 1}\,u_i = c\,\frac{u_i}{u_0} = c\,\frac{s_i}{s_0}
= c\,\frac{\psi^{\dagger}\,\tilde{\beta}_i\,\psi}{\psi^{\dagger}\,\psi}\,.
\label{eq:5}
\ee

Exactly the same procedure can be followed for massive bosons as for massive
fermions to determine the quantum potential and the Bohmian trajectories, except
that the sum over $a$ has to be carried out only over the independent degrees
of freedom (six for $\psi$ and six for $\bar{\psi}$ for spin-1 bosons).
The constraint (\ref{eq:4}) implies the four conditions $\vec{A} =
\vec{\nabla}\times\,\vec{B}$ and  $\vec{\nabla}\,.\,\vec{E} = 0$.

The theory of massless spin 0 and spin 1 bosons cannot be obtained simply by
taking the limit $m_0$ going to zero. One has to start with the equation
\cite{HC}

\be
i\,\hbar\,\beta_\mu \partial^\mu\,\psi + m_0\,c\,\Gamma\,\psi = 0
\label{eq:8}
\ee
where $\Gamma$ is a matrix that satisfies the following conditions:

\ben
\Gamma^2 &=& \Gamma\,\\
\Gamma\,\beta_\mu + \beta_\mu\,\Gamma &=& \beta_\mu\,.
\label{eq:9}
\een
Multiplying (\ref{eq:8}) from the left by $1 - \Gamma$, one obtains

\be
\beta_\mu\,\partial^\mu\, (\,\Gamma\,\psi\,) = 0\,.
\label{eq:10}
\ee
Multiplying (\ref{eq:8}) from the left by $\partial_{\lambda}\,
\beta^{\lambda}\,\beta^{\nu}$, one also obtains

\be
\partial^{\lambda}\,\beta_{\lambda}\,\beta_\nu\,(\,\Gamma\,\psi\,) =
\partial_\nu\, (\,\Gamma\,\psi\,)\,.
\label{eq:11}
\ee
It follows from (\ref{eq:10}) and (\ref{eq:11}) that

\be
\Box\,\, (\,\Gamma\,\psi\,) = 0
\label{eq:12}
\ee
which shows that $\Gamma\,\psi$ describes massless bosons. The Schr\"{o}dinger
form of the equation

\be
i\,\hbar\,\frac{\partial\, (\,\Gamma\,\psi\,)}{d t} = - i\,\hbar\,c
\tilde{\beta}_i\,\partial_i\, (\Gamma\,\psi)
\label{eq:13}
\ee
and the associated first class constraint

\be
i\,\hbar\,\beta_i\,\beta_0^2\,\,\partial_i\,\psi
+ m_0\,c\,(\,1 - \beta_0^2\,)\,\Gamma\,\psi = 0
\label{eq:14}
\ee
follow by multiplying (\ref{eq:8}) by $\beta_0$ and $1 - \beta_0^2$
respectively. The rest of the arguments are analogous to the massive case.
For example, the Bohmian 3-velocity $v_i$ for massless bosons can be defined by
equation (\ref{eq:5}).

Neutral massless spin-1 bosons have a special significance in physics. Their
wavefunction is real, and so their charge current $j_\mu =
\phi^{T}\,\beta_\mu\,\phi$ vanishes. However,
their probability current density $s_\mu$ does not vanish.
Furthermore, $s_i$ turns out to be proportional to the Poynting vector, as it
should.

Modifications to these equations can be introduced as
in the massive case to obtain a classical theory of massless bosons.

\section{The Gravitational Field}

Exactly the same procedure can also be applied to the gravitational field described
by Einstein's equations

\be
R_{\mu\nu} - \frac{1}{2}\,g_{\mu\nu}\,R = 0
\label{eq:a'}
\ee
for the vacuum, where $R_{\mu\nu}$ is the Ricci tensor and $R$ the curvature
scalar. In this section, following \cite{WD}, we will use the
signature $- + + +$ and the absolute system of units $\hbar = c = 16\, \pi\, G
= 1$. The decompostion of the metric is given by \cite{Holland}

\ben
ds^2 &=& g_{\mu\nu}\,d x^\mu\,dx^\nu\nonumber\\
&=& (\,N_i\,N^i - N^2\, )\,d t^2 + 2\,N_i\,dx^i\,d t + g_{ij}\,d x^i\,dx^j
\een
with $g_{i\,j}({\bf x})$, the 3-metric of a 3-surface embedded in space-time,
evolving dynamically in superspace, the space of all 3-geometries.

By quantizing the Hamiltonian constraint, one obtains in the standard
fashion the Wheeler-DeWitt equation \cite{WD}

\be
\left[\,G_{i\, j\, k\, l}\,\frac{\delta^2}{\delta g_{i\,j}\,\delta g_{k\,l}}
+ \sqrt{g}\,\,\,^3R\,\right]\,\Psi = 0
\label{eq:b'}
\ee
where $g =$ det $g_{i\,j}$, $^3R$ is the intrinsic curvature, $G_{i\,j\,k\,l}$
is the supermetric, and $\Psi [g_{i\,j}(x)]$ is a wave functional in
superspace. Substituting $\Psi = A\,\, exp\,\,(i\, S)$, one obtains as usual a
conservation law

\be
G_{i\,j\,k\,l}\,\frac{\delta}{\delta g_{i\,j}}\,\left(\,A^2\,\frac{\delta S}
{\delta g_{k\,l}}\,\right) = 0
\ee
and a modified Einstein-Hamilton-Jacobi equation

\be
G_{i\,j\,k\,l}\,\frac{\delta S}{\delta g_{i\,j}}\,\frac{\delta S}
{\delta g_{k\,l}} - \sqrt{g}\,\,\,^3R + Q = 0
\ee
where

\be
Q =  - A^{- 1}\,G_{i\,j\,k\,l}\,\delta^2\,A/\delta g_{i\,j}\,\delta g_{k\,l}
\label{eq:Z}
\ee
is the quantum potential. It is invariant under 3-space diffeomorphisms.
The causal interpretation of this {\it field} theory (as distinct from particle
mechanics considered earlier) assumes that the universe whose quantum state
is governed by equation (\ref{eq:b'}) has a definite 3-geometry at each
instant, described by the 3-metric
$g_{ij}({\bf x}, t)$ which evolves according to the classical Hamilton-Jacobi
equation

\be
\frac{\partial g_{i\,j}({\bf x}, t)}{\partial t} = \partial_i N_j
+ \partial_j N_i + 2\,N\,G_{i\,j\,k\,l}\,\frac{\delta S}
{\delta g_{k\,l}}\vert_{g_{i\,j}(x) = g_{i\,j}({\bf x},t)}
\label{eq:c'}
\ee
but with the action $S$ as a phase of the quantum wave functional.
This equation can be solved if the initial data $g_{i\,j}({\bf x}, 0)$
are specified. The metric in this field theory clearly corresponds to the
position in particle mechanics, equation (\ref{eq:c'}) being its guidance
condition.

It is now clear that one can modify the Wheeler-DeWitt equation (\ref{eq:b'})
to the form

\be
\left[\,G_{i\, j\, k\, l}\,\frac{\delta^2}{\delta g_{i\,j}\,\delta g_{k\,l}}
+ \sqrt{g}\,\,\,^3R - Q_{cl}\,\right]\,\Psi_{cl} = 0
\ee
where $Q_{cl}$ is defined by an expression analogous to (\ref{eq:Z}) with $A$ and $S$ replaced by the classical variables $A_{cl}$ and $S_{cl}$. This leads to the classical Einstein-Hamilton-Jacobi equation

\be
G_{i\,j\,k\,l}\,\frac{\delta S_{cl}}{\delta g_{i\,j}}\,\frac{\delta S_{cl}}
{\delta g_{k\,l}} - \sqrt{g}\,\,\,^3R = 0\,.
\ee
The term $Q_{cl}$ can then be interpreted, as before, as a potential arising due to
the coupling of gravitation with other forms of energy. If this coupling could
be switched off, quantum gravity effects would become important.
The question arises as to whether this can at all be done for gravitation.

\section{Concluding Remarks}

It is usually assumed that a classical system is in some sense a limiting
case of a more fundamental quantum substratum, but no general
demonstration for ensembles of systems has yet been given. That a quantum
system may, on the other hand, be a part of a classical system in which its
typical quantum features lie dormant is, however, clear from the above
discussions. The part therefore naturally shares the ontology of the total
classical system, and {\it the measurement problem does not even arise}. The
nonlocal quantum potential that is responsible for self-organization
and the creation of varied stable and metastable quantum structures, becomes
active only when the coupling of the part to the whole is switched off.
This is a clearly defined physical process that links the classical and
quantum domains.

According to this view, therefore, every quantum
system is a closed system and every classical system is an open system. The
first Newtonian law of motion therefore acquires a new interpretation---the
law of inertia holds for a system not when it is isolated from
everything else but when it interacts with
its environment to an extent that all its quantum aspects are quenched.
Various attempts to show that the classical limit of quantum systems is
obtained in certain limits, like large quantum numbers and/or large numbers
of constituents, have so far failed \cite{Mermin}. The reason is clear---a 
linear
equation like the Schr\"{o}dinger equation can never describe
a classical system which is described by 
a {\it modified} Schr\"{o}dinger equation with a nonlinear term. This
nonlinear term must be generated through some mechanism like the coupling of the system to 
 its environment.
There are, of course, other purely {\it formal} limits too (like $\hbar$ going 
to
zero, for example) in which a closed quantum system reduces to a classical
system, as widely discussed in the literature.

It is clear from the usual `decoherence' approach that the interaction of a quantum system with its environment in the form of some kind of heat bath is necessary to obtain a quasi-classical limit of quantum mechanics. This is usually considered to be a major advance in recent years. Such decoherence effects have already been measured in cavity QED experiments. Decoherence effects are very important to take into account in other critical experiments too, like the use of SQUIDs to demonstrate the existence of Schr\"{o}dinger cat states. The failure to observe cat states so far in such experiments shows how real these effects are and how difficult it is to eliminate them even for mesoscopic systems. I have taken these advances in our knowledge seriously in a phenomenological sense and tried to incorporate them into a conceptually consistent scheme. 

The usual decoherence approach however suffers from the following difficulty: it does neither solve the measurement problem nor does it lead to a truly classical phase space. The two problems seem to be intimately related. The density matrix becomes diagonal only in the coordinate representation. In other words, it does not represent a proper `statistical mixture'. The use of the linear Schr\"{o}dinger equation then automatically implies that the momentum space representation is necessarily non-diagonal. This does not happen in the approach advocated in this paper because of equation (\ref{eq:h1}) which guarantees the emergence of classical phase space and a proper `statistical mixture'. A clear empirical difference must therefore exist between the predictions of the usual decoherence approach and the approach advocated in this paper in the classical limit. It should be possible to test this by suitable experiments which are under consideration. The proposed conceptual frame!
work is therefore falsifiable.

\section{Acknowledgements}

The basic idea that quantum mechanics may be contained within classical
mechanics occurred during a discussion with C. S. Unnikrishnan in the
context of an experiment carried out by R. K. Varma and his colleagues
and Varma's theoretical explanation of it in terms of the Liouville equation
written as a series of Schr\"{o}dinger-like equations.
Unnikrishnan should not, however, be held responsible for any defects
of my particular formulation of the idea.

The first version of the non-relativistic part of this paper was presented at the ``Werner Heisenberg
Colloquium: The Quest for Unity: Perspectives in Physics and Philosophy''
organized by the Max Muller Bhavan, New Delhi, the National
Institute of Science, Technology and \& Development Studies, CSIR, Government
of India and the Indian Institute of Advanced Study, Shimla from 4th to
7th August, 1997 at Shimla.

The author is grateful to the Department of Science and Technology, Government of India, for a research grant that enabled this work to be completed.

\end{document}